\documentclass[aps, prb, twocolumn, showpacs, superscriptaddress, floatfix, 10pt]{revtex4-1}
\usepackage{framed}
\usepackage{graphicx}
\usepackage{amsmath}
\usepackage{epsfig}
\usepackage{helvet}
\usepackage{amssymb}
\usepackage{amsfonts}
\usepackage{bbm}
\usepackage{framed}
\usepackage{xcolor}
\usepackage{color}
\usepackage[colorlinks,bookmarks=false,citecolor=blue,linkcolor=red,urlcolor=blue]{hyperref}

\begin{document}

\title{Conversion of projected entangled pair states into a canonical form}
\author{Reza Haghshenas}
\author{Matthew J. O'Rourke}
\author{Garnet Kin-Lic Chan}
\affiliation{Division of Chemistry and Chemical Engineering, California Institute of Technology, Pasadena, California 91125, USA}

\begin{abstract}
  We propose an algorithm to convert a projected entangled pair state (PEPS) into a canonical form, analogous to the well-known canonical form of a matrix product state. Our approach is based on a variational gauging ansatz for the QR tensor decomposition of PEPS columns into a matrix product operator and a finite depth circuit of unitaries and isometries. We describe a practical initialization scheme that leads to rapid convergence in the QR optimization. We explore the performance and stability of the variational gauging algorithm in norm calculations for the transverse-field Ising and Heisenberg models on a square lattice. We also demonstrate energy optimization within the PEPS canonical form for the transverse-field Ising and Heisenberg models. We expect this canonical form to open up improved analytical and numerical approaches for PEPS.
\end{abstract}
\maketitle

%%%%%%%%%%% 
\section{Introduction}
Tensor network states (TNS)~\cite{White:1992,white1993density,vidal2007mera,
changlani2009approximating,mezzacapo2009ground} are widely used as variational wave functions to approximate low-energy states of quantum many-body systems~\cite{Verstraete:2008, Orus:2014}. Their power arises from their ability to efficiently capture global behaviors of quantum correlations in the system, as described by entanglement area laws~\cite{Vidal:2003, Calabrese:2004}. As a consequence, the global wave function is encoded in local tensors with finite bond dimension. A concrete example is the matrix product state (MPS)~\cite{
fannes1992finitely, ostlund1995thermodynamic, fannes1994, Schollwock:2011}, a class of tensor-network states that capture the area law in 1D, and which underlie the success of the density matrix renormalization group (DMRG)~\cite{White:1992, white1993density}.

The local tensors in a TNS are not uniquely defined and contain redundant parameters known as a local gauge.
In MPS, such gauges can be fixed by bringing the MPS into a canonical form where all tensors but one are
isometric \cite{Schollwock:2011}. The canonical form is simple to compute through QR decompositions, and has many applications, such as in defining optimal local truncations~\cite{Schollw:2005, Schollwock:2011}, the DMRG algorithm, constructing the tangent space of excitations~\cite{dorando2009analytic, haegeman2011time, nakatani2014linear, Vanderstraeten:2015}, and providing a framework to characterize phases~\cite{Pollmann:2012, Huang:2013}.

Projected entangled pair states (PEPS) \cite{nishino1996corner,Verstraete:2004,
verstraete2006criticality,Orus:2014} are higher-dimensional generalizations of MPS with analogous area laws. The PEPS has widely been used as a variational ansatz to explore physical properties of quantum many-body systems\cite{Murg:2009, Wang:2016, zheng2017stripe, Liu:2018,Haghshenas:2018May,  Haghshenas:2018, Matthew:2018, Haghshenas:2019}. It has already been observed that partially fixing the gauge of local tensors can dramatically improve the efficiency and stability of PEPS algorithms \cite{Lubasch:2014, Phien:2015, Phien:2015Jul, Evenbly:2018}. However, unlike in MPS, computing a fully canonical form for a PEPS remains a challenge. 

Here, we introduce a gauging variational ansatz that efficiently brings a PEPS wave function into a full canonical form in direct analogy with that of an MPS. To do so, we re-express the columns of the PEPS as a QR tensor product, where $Q$ is an isometric column tensor network and $R$ is a matrix product operator (MPO). We show that $Q$ can be compactly parametrized by a finite-depth circuit of block isometries and unitaries that can be determined by variational optimization. After transforming all columns but one (a central column) to be isometric, we obtain the (column) canonical form of the PEPS, where part of the entanglement in the environment is transferred to the central column. We explore the stability and performance of the QR decomposition and PEPS canonical representation in calculating the norm in the 2D transverse-field Ising (ITF) and Heisenberg models on a square lattice. We also analyze the behavior of imaginary-time energy optimization in the canonical PEPS form in the context of the ground-state of the ITF and Heisenberg models.

The paper is organized as follows. The basic concepts of the PEPS ansatz are introduced in Sec.~\ref{Sec:definition}. We first discuss the canonicalization procedure in the context of MPS in Sec.~\ref{Sec:Can} as a basis to describe our approach to addressing this problem for PEPS. We then study the cost, accuracy, and stability of our gauging variational ansatz in calculations on the ITF and Heisenberg models. We also compare the results of direct energy optimization in the canonical form to results from standard PEPS optimization algorithms. Finally, we summarize our findings in Sec.~\ref{Sec:Conclusion} and discuss future research directions.

\section{PEPS definition and background}
\label{Sec:definition}

\begin{figure}
\begin{center}
\includegraphics[width=1.0 \linewidth]{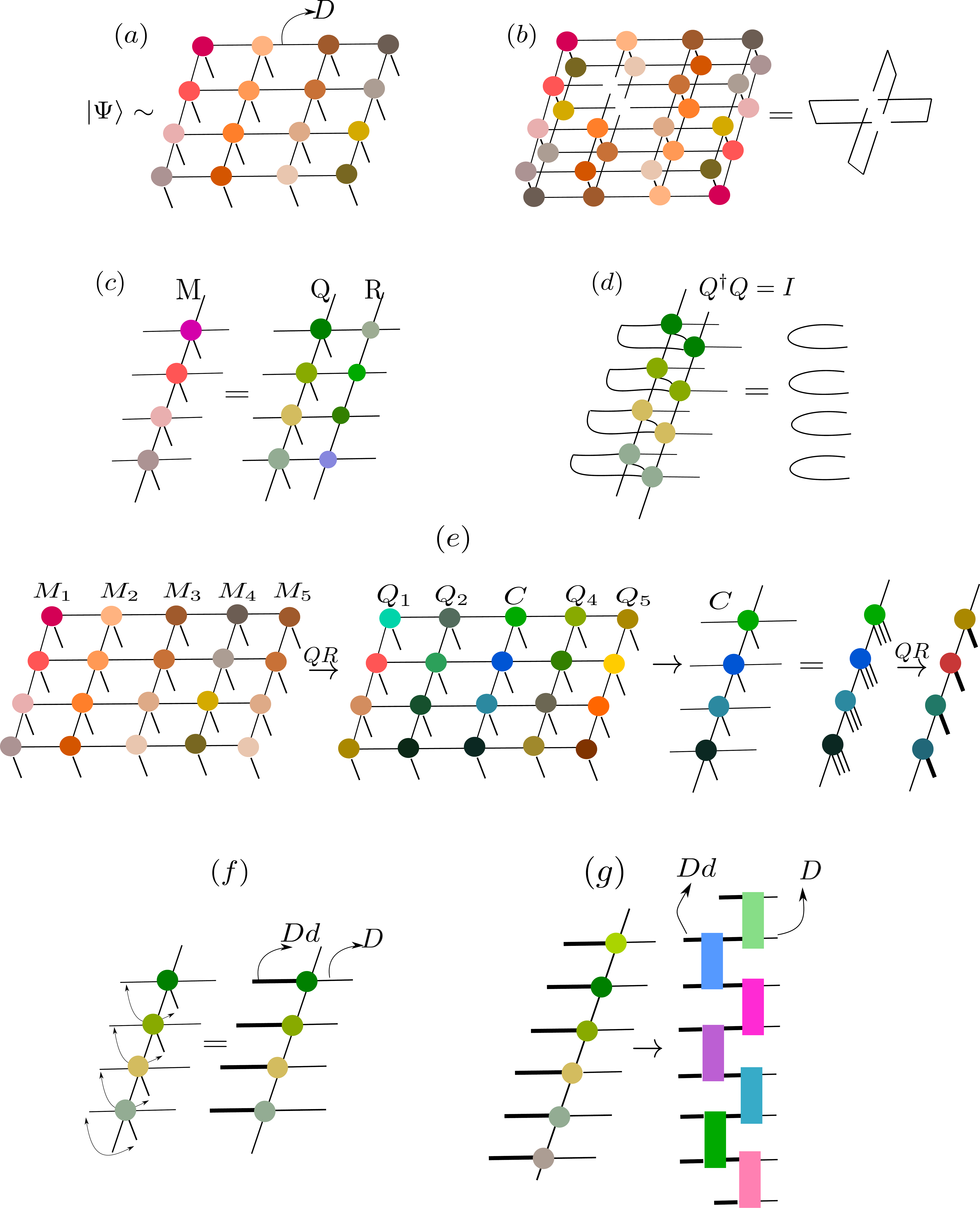} 
\caption{(Color online) (a) Tensor network diagram of the PEPS $|\Psi \rangle$ on a $4\times 4$ square lattice with open boundary conditions. Different colors are used for each tensor to explicitly indicate a non-translationally invariant tensor network. (b) A PEPS canonicalized around a single site, showing that the environment around that site contracts to the identity tensor. (c) A decomposition of a bulk column $M \approx QR$, where (d) the tensor network $Q$ is isometric, i.e. $Q^\dag Q=\mathbbm{1}$. (e) A graphical representation of the steps based on our QR scheme to bring a PEPS into canonical form. Note in the final step, MPS canonicalization is used on the central column $C$. (f) The tensor network $Q$ is reshaped into an MPO by fusing the virtual bond and physical bond, as shown by the arrows. The thick virtual bonds have bond dimension $Dd$. (g) The isometric tensor network $Q$ is parameterized by a finite depth circuit of $l$-site isometries/unitaries $\{ u_i\}$.} 
  \label{fig:peps}
\end{center}
\end{figure}

A PEPS is a TNS defined by a set of local tensors $\{ A^{s_i}_{i} \}$ connected by virtual bonds along the grid of the physical lattice. The bond dimension of the virtual bonds is denoted $D$, which controls the number of parameters (or, more physically, the amount of entanglement in the wavefunction) and hence the accuracy of the ansatz. The physical indices $s_i$ encode the local physical Hilbert space of dimension $d$. A PEPS wave function $|\Psi \rangle$ on the $l_x \times l_y=4 \times 4$ square lattice with open boundary conditions is depicted in Fig.~\ref{fig:peps}(a),
\begin{align}
  |\Psi \rangle = \sum_{\{ s_i\}}   \mathcal{F}(A^{s_1}_1, A^{s_2}_2, \cdots, A^{s_{l_x \times l_y}}_{l_x \times l_y}) |s_1, s_2 \cdots, s_{l_x \times l_y} \rangle
  \label{eq:peps}
\end{align}
where $\mathcal{F}$ denotes tensor contraction of the virtual bonds. The tensors are all colored differently to indicate that we do not assume translational invariance in the tensor network.

The tensor contraction in Eq.~\eqref{eq:peps} is invariant under insertion of a gauge matrix and its inverse $G$, $G^{-1}$ between two tensors (along with a virtual bond). In an MPS, the canonical form at site $i$ is defined as the choice of gauges such that the environment tensor $\mathcal{G}_i$ (constructed by partial norm-contraction over all sites except $i$) is the identity tensor
\begin{align}
  \mathcal{G}_i = \mathcal{F} \left(\prod_{j\neq i} {E}_j\right) = \mathbbm{1}, \label{eq:canonical}
  \end{align}
with $E_j = \sum_{s_j} A_j^{s_j \dagger} A_j^{s_j}$; $\mathbbm{1}$ denotes the tensor $\delta_{i_1i_1'}\delta_{i_2i_2'}\ldots$ where $i_1i_2\ldots$, $i_1'i_2'\ldots$ index
the virtual bonds of $A^{s_i\dagger}_i$, $A^{s_i}_i$ respectively.
By ensuring that the PEPS tensors satisfy Eq.~\eqref{eq:canonical}, it also defines an analogous canonical form for a PEPS, depicted in Fig.~\ref{fig:peps}(b). In the case of an MPS, we can convert an arbitrary MPS into canonical form by sequential QR (LQ) decompositions of tensors to the left (right) of site $i$, $A_j^{s_j} \to Q_j^{s_j} R_j$ ($A_j^{s_j} \to L_j Q_j^{s_j}$) where $Q_j$ is orthogonal in the sense $\sum_{s_j} {Q_j^{s_j\dag}} Q_j^{s_j} = \mathbbm{1}$ (for LQ, $\sum_{s_j} {Q_j^{s_j}} {Q_j^{s_j\dag}} = \mathbbm{1}$). For simplicity, we henceforth do not distinguish between QR and LQ, with the choice implicit from the diagrammatic representation. $R_j$ is then absorbed into the adjacent tensor for the subsequent QR decomposition until the full canonical form is reached~\cite{Schollwock:2011}.

%%%%%%%%%%%%%%%%%%%%%%%%%%%%%%%%%%Fig. 1%%%%%%%%%%%%%%%%%%%%%%%%%%%%%%%%%%%%%%%%%%
 %%%%%%%%%%%%%%%%%%%%%%%%%%%%%%%

\section{PEPS canonical form and column QR ansatz}
\label{Sec:Can}

To similarly canonicalize a PEPS, we sequentially decompose the PEPS columns, denoted $M$ (composed of tensors $\{ m_i \}$), as a QR tensor contraction, where the column tensor network $Q$ is isometric, satisfying $Q^\dag Q=\mathbbm{1}$,  see Fig.~\ref{fig:peps}(c, d). The gauge column tensor network $R$ (composed of tensors $\{ r_i\}$) is an MPO acting on the horizontal virtual bonds. Once all columns (around a central column) are decomposed to be isometric, the central column $C$ can be viewed as an MPS by grouping the horizontal bonds with the physical bonds. This central column can then be canonicalized around a chosen site using the MPS canonicalization algorithm above, to yield a complete PEPS canonicalization (Fig.~\ref{fig:peps}(e)). Note that the PEPS canonicalization condition around a site (Fig.~\ref{fig:peps}(b)) does not itself specify that columns to the left and right of the central column separately contract to the identity; the conditions we impose are thus sufficient and convenient when canonicalizing a PEPS, but are more constrained than the necessary conditions for Fig.~\ref{fig:peps}(b). 

To explicitly carry out the QR decomposition, we first rewrite $M$ and thus $Q$ as MPOs by fusing physical bonds with the left virtual bonds (Fig.~\ref{fig:peps}(f))---for the equivalent LQ decomposition, the physical bonds should be fused with the right virtual bonds, as in MPS. Then, to explicitly enforce the isometric constraint on $Q$, we write it as a finite depth-$n$ circuit of block-size $l$ isometries and unitaries $\{ u_i\}$, where the isometries appear in the edge layer of the circuit (Fig.~\ref{fig:peps}(g): the isometries have thick bonds (dimension $Dd$) and thin bonds (dimension $D$) while the unitaries only have thick bonds (dimension $Dd$))\cite{Pollmann:2016, Wahl:2017}. The tensors in the first layer are chosen to be unitary and those in the remaining layers are isometries. The layer depth and block size control the distribution of entanglement between $Q$ and $R$. In practice, to obtain a faithful QR decomposition we have found it sufficient to use $n=2$ (a single layer of unitaries and isometries), increasing $l$ if necessary. In addition, we set the vertical bond dimension of $R$ equal to that of $M$.

\begin{figure}
\begin{center}
\includegraphics[width=1.0 \linewidth]{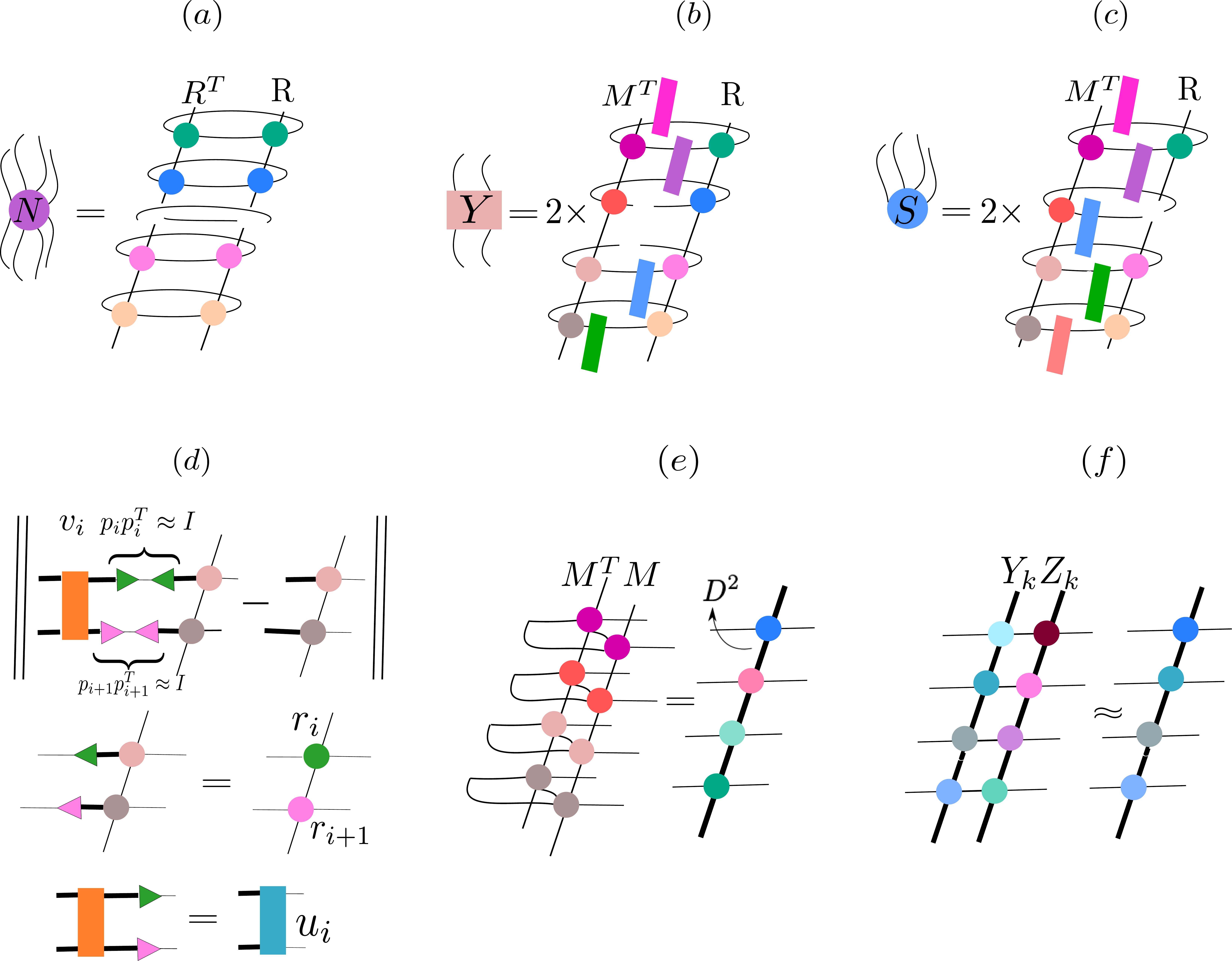} 
\caption{(Color online) (a, b, c) Tensor network representation of the tensors $N$, $S$ and $Y$ appearing in the QR  optimization procedure; free indices in the first column correspond to the left indices
of the tensors in matrix form, free indices in the second column correspond to the right indices of the tensors in matrix form. (d) The local distance (cost function) used to obtain a good initial guess for local tensors $u_i$, $r_i$ and $r_{i+1}$. The cost function is minimized with respect to tensors $p_i$, $p_{i+1}$ and $v_i$, which are used to build $u_i$, $r_i$ and $r_{i+1}$ as depicted. (e, f) Graphical representation of $M^\dag M$ and $Y_kZ_k$ used in the Schulz algorithm.} 
  \label{fig:peps1}
\end{center}
\end{figure}

To determine the tensors in the QR ansatz, we minimize the distance (cost function) ${F} = \parallel M-Q(\{ u_i\})R(\{r_i\})\parallel$ with respect to $\{u_i, r_i\}$  ($\parallel \cdot \parallel$ is the Hilbert-Schmidt norm) using standard tensor network techniques~\cite{Evenbly:2009, Kao:2015, Evenbly:2017}. We optimize the tensors one at a time and sweep until convergence. The cost function depends quadratically on $\{r_i\}$, i.e. ${F}= r_i^{\dagger}Nr_i-Sr_i+\textrm{const}$, which is explicitly minimized by solving the linear equation $Nr'_i=S$.
%% This can be solved iteratively with a method such as conjugate gradients, which only requires matrix-vector multiplication at a cost of $\mathcal{O}(D^8)$.
To update the isometric/unitary tensors $\{u_i\}$, we observe that the cost function only depends linearly on them due to cancellations, i.e. ${F}= u_i^{\dagger}Y+\text{const}$, thus the optimal solution is given by $u'_{i}=-VU^{\dagger}$, where $V$, $U$ appear in the singular value decomposition $Y=UsV^{\dagger}$~\cite{Evenbly:2009}.
%% {\color{red}Computing the cost of $U$ and $V$ by minimization
%%   costs [...], while using an explicit
%%   SVD construction has a cost of $\mathcal{O}(D^{12})$. -- Should we discuss the costs only later?}
The tensor network diagrams of $N$, $S$ and $Y$ appear in Fig.~\ref{fig:peps1}(a, b, c) respectively.

%%%%%%%%%%%%%%%%%%%%%%%%%%%%%%%%%%Fig. 2%%%%%%%%%%%%%%%%%%%%%%%%%%%%%%%%%%%%%%%%%%
 %%%%%%%%%%%%%%%%%%%%%%%%%%%%%%%

To accelerate the QR optimization (and to avoid local minima) we start with a good initial guess for $\{u_i, r_i\}$.
We have used two techniques. The first uses a local projective truncation on the tensors $\{m_i\}$ to initialize
$\{ u_i\}$ and $\{ r_i\}$. To this end, we contract an approximate resolution of the identity (i.e. $\mathbbm{1}_{D}\approx p_ip_{i}^{\dag}$ where $p_i$ is a local isometry) and a unitary $v_i$ into two adjacent tensors $m_i, m_{i+1}$ and optimize $p_i$, $p_{i+1}$ and $v_i$ to minimize the local cost function shown in Fig.~\ref{fig:peps1}(d). Once we have the optimized tensors $p_i, p_{i+1}$ and $v_i$, we construct a guess for $r_i, r_{i+1}$ and $u_i$ as shown in  Fig.~\ref{fig:peps1}(d). This initialization is purely local but in practice, we find that it performs well.

A second strategy is based on an accurate estimate of the $\{ r_i \}$ tensors using a Schulz iteration for the matrix square-root \cite{Higham:1997}. Note that $R$ is formally the square root of $M^\dag M$, due to the isometric property of $Q$. We thus rewrite $M^\dag M$ as an MPO as in Fig.~\ref{fig:peps1}(e). Then, starting from $Y_0=M^\dag M$ and $Z_0=\mathbbm{1}$, the coupled Schulz iteration, $Y_{k+1}=\frac{1}{2} Y_k(3\mathbbm{1}-Z_kY_k)$, $Z_{k+1}=\frac{1}{2} (3\mathbbm{1}-Z_kY_k)Z_k$, gives $Y_{k\rightarrow \infty}=R$ and $Z_{k\rightarrow \infty}=R^{-1}$. The vertical bond dimension of $Y_k$ and $Z_k$ increases with each MPO multiplication (Fig.~\ref{fig:peps1}(f)) thus we perform MPO compression after each iteration (viewing the MPO as an MPS). The vertical bond dimension of the final $Y_k$ ($R$) is compressed back to the original bond dimension of $M$. Also, since $Z_k$ approximates $R^{-1}$ which may have arbitrarily large norm, we regularize the iteration using $M^\dag M\rightarrow M^\dag M +\delta I$, where $\delta$ is a small number ($\sim 10^{-6}$). The Schulz iteration converges rapidly (see SM) and we use this accurately estimated  $R$ to initialize the optimization of the tensors in $Q$ with respect to the cost function $F$. Although computing the Schulz iteration is more expensive than the local initialization, we expect it to be better when canonicalizing PEPS with more entanglement.

\subsection{PEPS canonicalization sweep and truncations}
\label{Sec:methodology}

To canonicalize all columns $M^{[1]} M^{[2]} \ldots M^{[l_x]}$ in the PEPS, we sweep over all the columns in a prescribed order (say from left to right) and compute the QR decomposition to each.
%% For a single PEPS column of
%% bond dimension $D$ and an $n=2$, $l=2$ QR ansatz, {\color{blue}the cost of the QR optimization is $\mathcal{O}(l_yD^{8})$} (using the iterative cost of solving for $\{ r_i \}$).
After  column $M^{[1]}$ has been converted to QR form, we then absorb the $R$ gauge into the neighboring $M^{[2]}$ column, creating a combined column $C^{[2]}$ with an increased vertical bond dimension of $D^2$. To avoid increasing the vertical bond dimension of
  subsequent columns, we compress $C^{[2]}$ to a smaller vertical bond dimension $D_c < D^2$. We can perform this column truncation as an MPS truncation with enlarged physical bond dimension $D^2d$.
  %% note that if the initial PEPS is already in canonical form and we
  %% are using QR to move the central column, then the surrounding columns are already canonicalized.
  The role played by $D_c$ is somewhat related to the $\chi$ in PEPS contraction algorithms \cite{Corboz:2010,Corboz:2010:April, Lubasch:2014}, but here $D_c$ is an (auxiliary) bond dimension for a single PEPS layer, rather than for a double layer. A more relevant comparison is therefore to the $\chi$ used in single-layer PEPS algorithms~\cite{Frank:2011, Xie:2017}, which is argued to be $\propto D$, thus leading us to conjecture that (asymptotically) $D_c \propto D$. Continuing, we perform the QR decomposition on the truncated $C^{[2]}$ column (with vertical bond dimension $D_c$), absorb $R$ into $M^{[3]}$ and proceed as before over the remaining columns, to finally
  produce a PEPS in the canonical form as $Q^{[1]} Q^{[2]} \ldots Q^{[l_{x-1}]} C^{[l_x]}$. Note that if MPS truncation is used
  to compress $C$, then the final central column will be in canonical form around a single site.

Canonicalization redistributes entanglement in the PEPS, thus the canonicalized PEPS has different
bond dimensions than the original PEPS. If we use the $n=2$, $l=2$ ansatz for $Q$, then when viewed as an MPO the $Q$ columns have a vertical bond dimension of $\mathcal{O}(D^2)$, while the central column $C$ has a vertical bond dimension of $D_c$. The formally large bond dimension of $Q$ is primarily an artifact of expressing the isometric constraint in terms of gates. Thus it is computationally most efficient to use the structure of $Q$ (i.e. viewing the column of isometries and column of unitaries separately) in the tensor network contractions.

Given a canonical PEPS, the canonicalization sweep can be used to convert between canonical forms (where we move the central column)
which is important in algorithms such as energy optimization. The only difference then from the canonicalization sweep discussed above
is that the $Q$ columns have a larger vertical bond dimension than $M$. Thus, when absorbing $R$ into a neighboring $Q$ column (in the $n=2$, $l=2$ ansatz), we create a central column $C$ of vertical bond dimension $D^2 D_c$, which we subsequently compress to $D_c$.

From the above, we see that in computing the canonical form, and in moving the central column, there are two potential sources of error that must be controlled. The first is the QR \emph{approximation error}, controlled by the finite-depth/block-size ($n$, $l$) of $\{ u_i\}$ and the vertical MPO bond dimension $D$ of $R$ (which we fix). The second is the \emph{absorption error}, that arises from the truncation of the central column's vertical bond dimension to $D_c$.

\subsection{Cost of conversion to canonical form}

We now discuss the leading costs of the computational steps in the conversion to the canonical form, assuming
the $n=2$, $l=2$ ansatz for Q and setting the vertical bond dimension of $R$ to always be equal to that of the column that is being decomposed.

\noindent {\it QR optimization}. For a column of vertical bond dimension $D$, the cost to determine the isometries/unitaries is $\mathcal{O}(l_yD^{6})$ (non-iterative) and $\mathcal{O}(l_yD^{4})$, using an iterative algorithm to solve the linear equation ${F}= u_i^{\dagger}Y+\text{const}$. The cost to determine $\{r_i\}$ tensors is $\mathcal{O}(l_yD^{12})$ (non-iterative) and $\mathcal{O}(l_yD^{8})$, using a minimization algorithm to solve ${F}= r_i^{\dagger}Nr_i-Sr_i+\textrm{const}$.

\noindent  {\it Absorption step}. As discussed above, when converting a standard PEPS with columns $M^{[1]} M^{[2]} \ldots M^{[l_x]}$ that have vertical bond dimension $D$ into canonical form, the typical absorption step during the sweep creates a central column $C$ with an enlarged vertical bond dimension $D D_c$ (because $R$ has dimension $D_c$). Compressing this down to a vertical bond dimension $D_c$, using sequential SVD on the MPS bonds, costs $\mathcal{O}(l_yD^{5}D_c^3)$. Alternatively, if we consider direct minimization $|| \, |\phi\rangle - |\psi\rangle ||$ (the cost typically reported in boundary PEPS algorithms) where $|\phi\rangle$ and $|\psi\rangle$ are MPSs with physical bond dimension $dD^2$ and virtual bond dimensions $D D_c$ and $D_c$ respectively, the cost is $\mathcal{O}(l_yD^{4}D_c^4)$.

Alternatively when carrying out the absorption step for a PEPS already in canonical form, e.g. when moving $C$ from $l_x$ to $l_{x-1}$ in the PEPS with columns $Q^{[1]} Q^{[2]} \ldots C^{[l_x]}$, then the absorption step involves compressing $C^{[l_{x-1}]}$ from vertical bond dimension $D^2D_c \to D_c$. Using sequential SVD on the MPS bonds, the cost is $\mathcal{O}(l_y D^8D^3_c)$, while direct minimization gives a cost of $\mathcal{O}(l_y D^{6}D^3_c)+\mathcal{O}(l_y D^{4}D^4_c)$). 

However, in this case, since $C^{[l_{x-1}]} = Q^{[l_{x-1}]} R$, we can use the ansatz structure of $Q$ to reduce the cost of the truncation, by absorbing and truncating first the column of isometries, then the column of unitaries. (In both these truncations, the surrounding columns are canonical, and thus each can be performed as an MPS truncation). With this technique, the cost of the absorption step is reduced to $\mathcal{O}(l_yD^4 D^4_c)$ by using direct minimization. The cost is the same for truncating the column of isometries and for the column of unitaries. Using sequential SVD truncation, the leading term is $\mathcal{O}(l_yD^{6}D^3_c)$.

\begin{figure}
\begin{center}
\includegraphics[width=1.0 \linewidth]{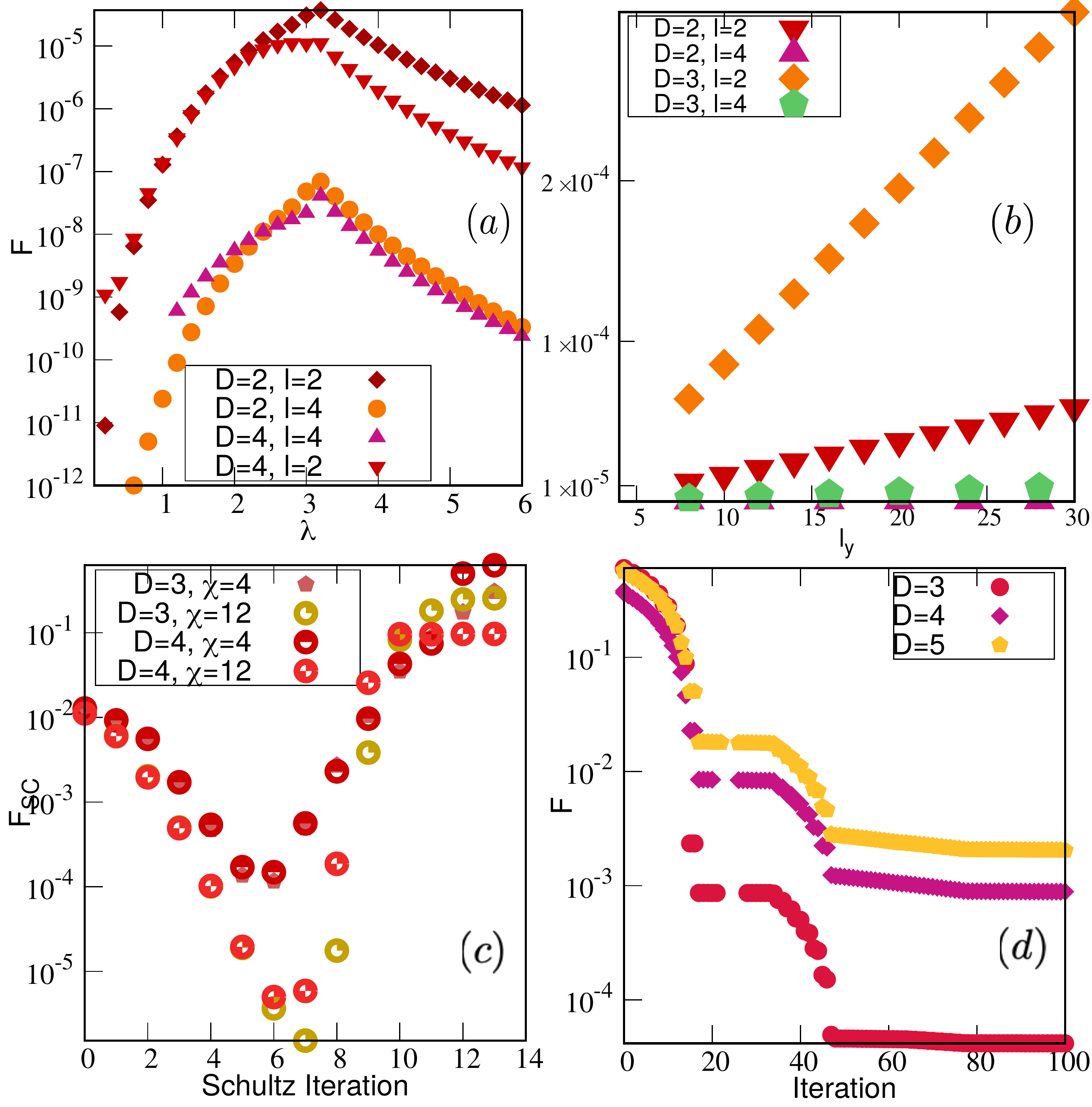} 
\caption{(Color online) Accuracy of the variational ansatz for QR decomposition of a single bulk
column. (a) The relative distance $F$ as a function of transverse Ising field $\lambda$, for a 
single bulk column with $l_y=16$, and for given values of bond dimension $D$ and block size $l$
used in $Q$. The distance $F$ rapidly decreases with increasing $l$, as more entanglement is included in the $Q$ circuit. (b) The relative distance as a function of PEPS column length $l_y$ for different  block sizes $l$.
  The error of the variational ansatz increases linearly with the length $l_y$. (c) The distance $F_{SC}$ as a function of Schulz iteration for a single bulk column with $l_y=16$ at magnetic field $\lambda=2.0$. (d) The relative distance $F$ of the Heisenberg model versus iteration number of the QR optimization for a single bulk column with $l_y=10$ and with different bond dimensions $D$.}  
  \label{fig:ACC}
\end{center}
\end{figure}
 %%%%%%%%%%%%%%%%%%%%%%%%%%%%%%%

%% Note that because the canonicalized PEPS has some bonds with different dimension to that in the original PEPS, care must be taken when comparing the cost of canonical and non-canonical PEPS algorithms. 

\subsection{Accuracy of QR ansatz}

To assess the accuracy of the QR ansatz, we first study its performance for a single PEPS column. As our initial state, we use the (approximate) ground-state of the spin-$\frac{1}{2}$ ITF and Heisenberg models on the square lattice. % (additional results for the Heisenberg model are in the SM).
The ITF model and the Heisenberg model are respectively defined by 
\begin{eqnarray*}
H_{\text{ITF}}=-\sum_{\langle ij\rangle} \sigma^i_z\sigma^j_z -\lambda \sum_i \sigma_x,\\
H_{\text{Heisenberg}}=\sum_{\langle i,j\rangle}\textbf{S}_{i}\cdot\textbf{S}_{j},\\
\end{eqnarray*} 
where ${\textbf{S}_i}\equiv(\sigma_x, \sigma_y, \sigma_z)$ and $\sigma_\alpha$ are the Pauli matrices. The ITF model has a critical point at $\lambda_c \approx 3.05$. Our initial PEPS is constructed from the bulk tensors of an infinite PEPS ground-state~\cite{Jordan:2008} (optimized with a full-update scheme~\cite{Corboz:2010:April, Phien:2015} and a $2\times 2$ unit cell) that is repeated periodically across the finite PEPS lattice.

We measure the accuracy of the QR ansatz by the value of its optimization cost function $F$. Here, the parameter controlling the accuracy is the block size $l$ of the isometric/unitary circuit (the number of layers is kept as $n=2$, and the vertical bond dimension of $R$ is kept as $D$). In Fig.~\ref{fig:ACC}(a), we show the plot of the distance $F$ versus ITF magnetic field $\lambda$. As expected, when the system is close to criticality, the accuracy is reduced as the ground state becomes more entangled. Increasing $l$ increases the disentangling effect of the unitaries, and the accuracy increases rapidly (we conjecture exponentially with $l$), especially far from criticality.

Next, we investigate the QR accuracy as a function of system size $l_y$ for the ITF model at field strength $\lambda=3.5$. As shown in Fig.~\ref{fig:ACC}(b),
the relative error in $F$ increases linearly with system size, although the slope shows a rapid decay with the isometry/unitary block size $l$. Thus, the variational gauging ansatz introduces a constant error per lattice site, consistent with a fidelity that goes like $e^{-\epsilon l_y} \sim 1-\epsilon l_y$.

%%%%%%%%%%%%%%%%%%%%%%%%%%%%%%%%%%Fig. 4%%%%%%%%%%%%%%%%%%%%%%%%%%%%%%%%%%%%%%%%%%
\begin{figure}
\begin{center}
\includegraphics[width=1.0 \linewidth]{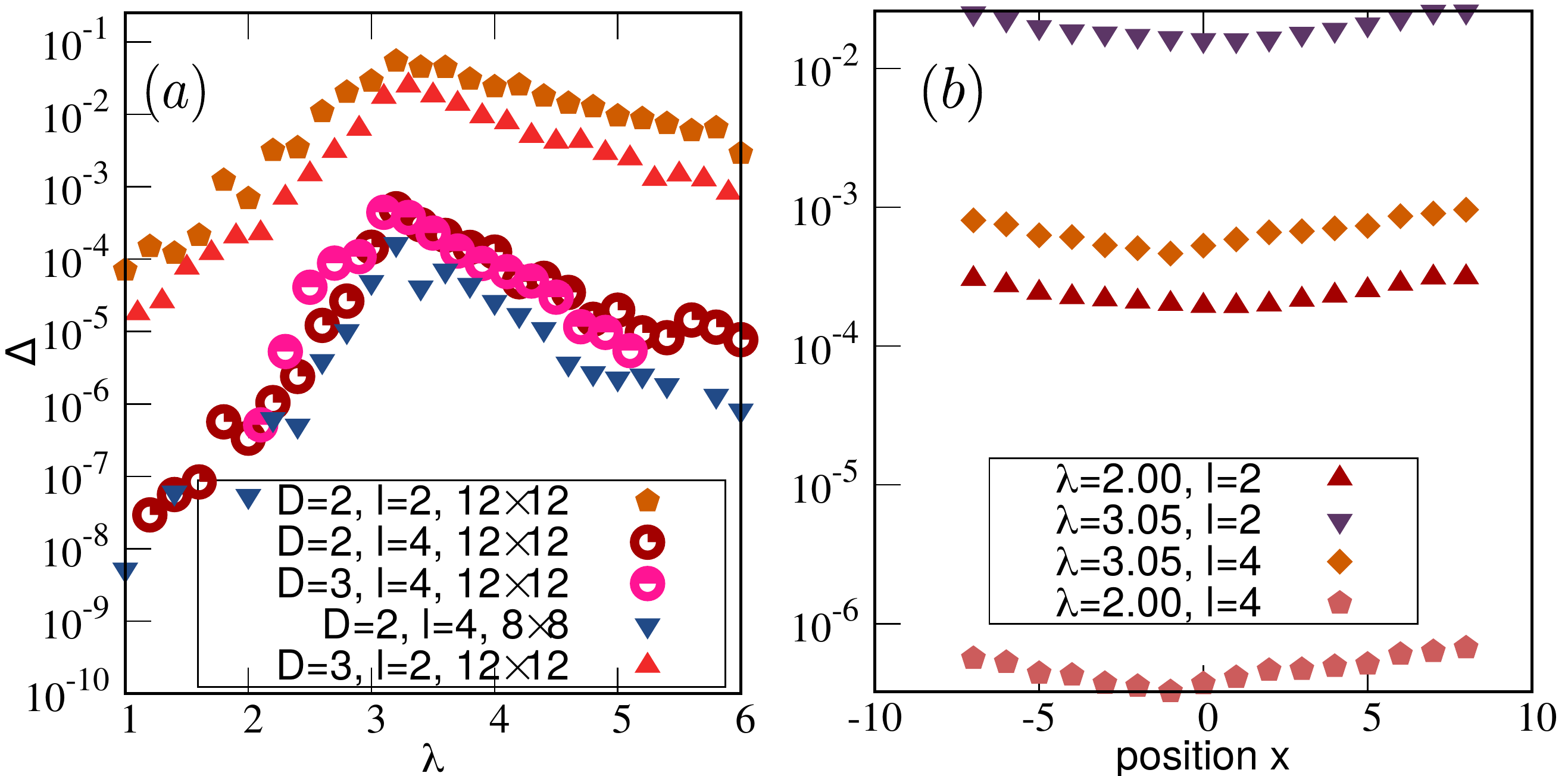} 
\caption{(Color online) (a) The relative error of the norm contraction $\Delta= \frac{N-N_{b}}{N_{b}}$ as a function of $\lambda$. $N_{b}$ is a reference norm obtained by boundary method contraction. For the norm computed via the canonical form, we use $D_c = 3D$. (b) $\Delta$ as a function of column position $x$ in the PEPS for different values of $\lambda$. The bond dimensions are $D=3$, $D_c=12$, and the lattice size is $16 \times 16$. }%{\color{red}(c) Log-linear plot of $\Delta$ as a function of successive canonicalization sweeps.}}
\label{fig:Norm}
\end{center}
\end{figure}
 %%%%%%%%%%%%%%%%%%%%%%%%%%%%%%%

We can also study the accuracy of the Schulz iteration. We show this by evaluating the distance $F_{SC}=\parallel M^\dag M - R^{2}  \parallel$ as a function of the Schulz iteration. The MPO compression is controlled by a truncated bond dimension $\chi$ (in the final iteration in the canonicalization algorithm, this is always set to $D$). In Fig.~\ref{fig:ACC}(c), we show how the accuracy of the Schulz iteration depends on $\chi$ for different initial bond dimensions $D$ for the ITF model at magnetic field $\lambda=2.0$. The regularization parameter is always set to $\delta\sim 10^{-6}$.    

Finally, we give additional results for the QR optimization in the Heisenberg model. We  plot the relative distance $F$ versus iteration number for a single bulk column with $l_y=10$ in Fig.~\ref{fig:ACC}(d), using block size $l=2$. We see that the relative error of the norm contraction $\Delta$ is similar to that of the ITF model at the critical point ($\sim 10^{-2}$) for a lattice of size $10\times10$.

\subsection{Accuracy of PEPS canonical form}

 We next investigate the accuracy and stability of the full PEPS canonical form constructed from a sweep of the QR approximation and absorption steps across the columns. We estimate the faithfulness of the canonical form from the norm contraction $\mathcal{N}=\langle\Psi | \Psi\rangle$. We compute the norm in the canonical form using only the central column $C$ since all other columns contract exactly to the identity. The relative error of norm contraction is then defined as $\Delta= \frac{\mathcal{N}-\mathcal{N}_{b}}{\mathcal{N}_{b}}$, where the reference value $\mathcal{N}_{b}$ is obtained using an accurate boundary contraction of the original (uncanonicalized) PEPS keeping a large boundary auxiliary bond dimension \cite{Verstraete:2004, Verstraete:2008, Lubasch:2014Aug}. In Figs.~\ref{fig:Norm}(a, b), we show a plot of the relative error $\Delta$ as a function of ITF magnetic field $\lambda$ (using the same approximate ground-state as above) and central column position. Similarly to the single-column results above, the accuracy of the full canonical form depends on the correlation length of the model, and the canonicalization error decreases rapidly (exponentially) as the block size $l$ is increased.

%%%%%%%%%%%%%%%%%%%%%%%%%%%%%%%%%%Fig. 5%%%%%%%%%%%%%%%%%%%%%%%%%%%%%%%%%%%%%%%%%%
\begin{figure}
\begin{center}
\includegraphics[width=1.0 \linewidth]{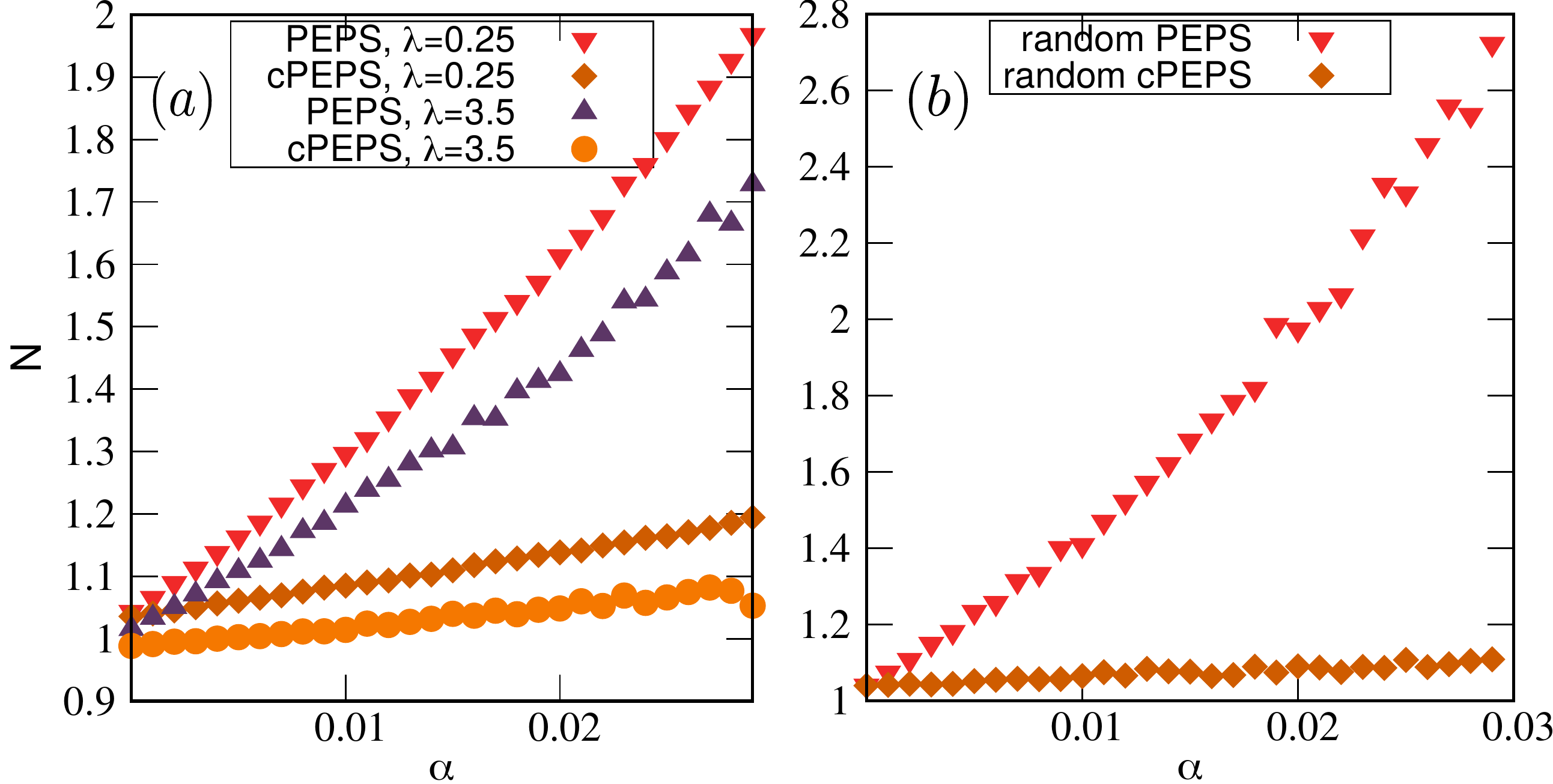} 
\caption{(Color online) The total norm of a $8 \times 8$ PEPS as a function of perturbation strength $\alpha$ for $(a)$ ITF ground-state PEPS at different coupling parameters $\lambda$ and for a uniform random PEPS. cPEPS denotes canonical PEPS derived from the standard PEPS. The bond dimensions are chosen to be $D=3$ and $D=2$ for the ITF PEPS ground state and random PEPS respectively.}
\label{fig:NormS}
\end{center}
\end{figure}
 %%%%%%%%%%%%%%%%%%%%%%%%%%%%%%%

%%%%%%%%%%%%%%%%%%%%%%%%%%%%%%%%%%Fig. 6%%%%%%%%%%%%%%%%%%%%%%%%%%%%%%%%%%%%%%%%%%
\begin{figure}
\begin{center}
\includegraphics[width=1.0 \linewidth]{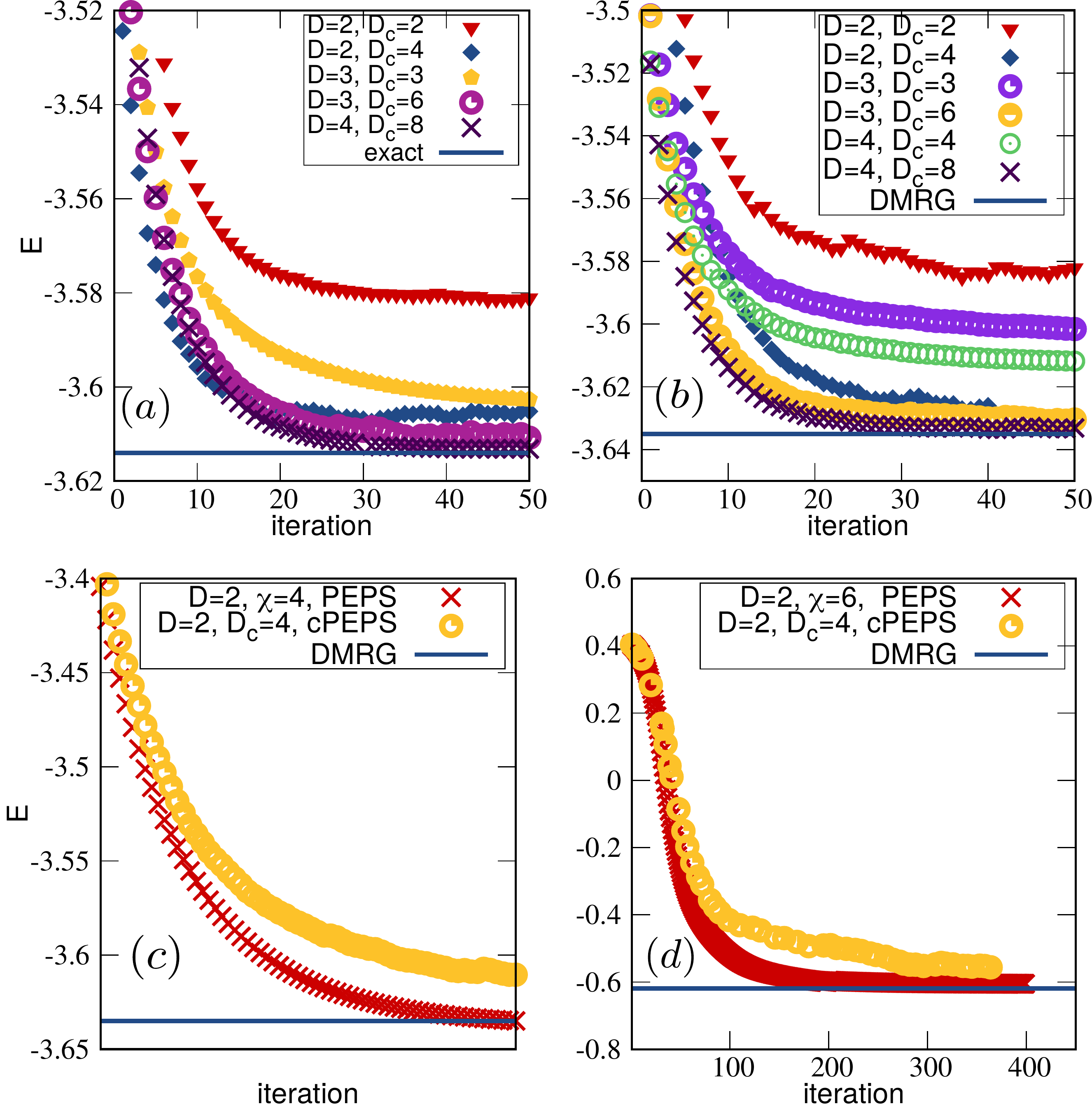} 
\caption{(Color online)  {Imaginary time energy optimization based on the PEPS canonical form. The canonical PEPS energy of the ITF model as a function of imaginary time sweep for (a) $4\times 4$ and (b) $8\times 8$ square lattices with field $\lambda = 3.5$. A direct variational energy comparison between canonicalized PEPS (cPEPS) and standard uncanonicalized PEPS\cite{Lubasch:2014Aug} for the (c) ITF model at $\lambda = 3.5$ and (d) Heisenberg model, on a $8\times 8$ square lattice. Both algorithms start form the same random initial PEPS. The horizontal DMRG line is the result of a converged 2D DMRG calculation and can be taken to be numerically exact. The parameter $\chi$ stands for boundary bond dimension in environment calculations in the standard PEPS optimization. cPEPS denotes a canonical PEPS.}}
\label{fig:Energy}
\end{center}
\end{figure}
 %%%%%%%%%%%%%%%%%%%%%%%%%%%%%%%

\subsection{Norm stability of the canonical form}
{An important property to assess is the numerical stability of the PEPS in the canonical form. We can study this by examining the stability of the total norm $N$ with respect to small perturbations of the tensors. A numerically stable form is one where small perturbations to the tensors result in small perturbations in the norm. To study this property, we compare the stability of the norm in a standard PEPS versus in its canonical form. We add small perturbations to each tensor  $A^i_{s_i} \rightarrow A^i_{s_i} + \alpha P$ in a canonical PEPS and a normal PEPS (the unperturbed canonical PEPS is obtained by first converting the normal PEPS to canonical form). The normalized tensors $P$ ($|P|=1$) are chosen to form a uniform random distribution in the interval $[0, 1]$. In Fig.~\ref{fig:NormS}(a, b), we plot the norm $N$ as a function of perturbation strength $\alpha$ for the ITF ground-state PEPS and a random PEPS. We observe, in both cases, that the canonical PEPS (cPEPS) remains more stable, as the norm changes less with the perturbation as compared to the standard PEPS. This is a result of the fact that the $Q$ tensors are isometric and adding small perturbations to them only affects the isometric property on the unit scale.}

\subsection{Energy optimization in canonical form}
\label{Sec:energy}

A natural application of the PEPS canonical form is to ground-state energy optimization, which mimics the use of the MPS canonical form in energy optimization. To show this, we perform imaginary time evolution on the ITF model, which we carry out with a sequence of gates $e^{-\tau h}$ on the horizontal and vertical bonds~\cite{Verstraete:2008, Lubasch:2014}. Evolution on a column of vertical bonds is conveniently carried out on the central column $C$ of a canonical PEPS with bond dimension $D_c$, where it reduces to an MPS imaginary time evolution followed by an MPS truncation with an enlarged physical bond dimension $D^2d$. %The imaginary time update and compression can be carried out with cost $\mathcal{O}(D^{2}D_c^{3})$ per bond. 
Evolution on a column of horizontal bonds can be carried out using a two-column canonical PEPS (analogous to the two-site MPS canonical form) where there are two central columns, and columns to the left and right of these two are isometric tensors $Q$, thus reducing
the optimization problem to one of a PEPS with only two columns. In this case, rather than canonicalizing the remaining environment around the bond in the two-column PEPS, we contract it exactly, which is straightforward. Since there are only two columns, these can be reduced to an MPS with enlarged vertical bond dimension $D^2D_c$ and physical bond dimension $D^2d^2$. %The cost to do an update and compression would be $\mathcal{O}(D^{7}D_c^{3})+\mathcal{O}(D^{8}D_c^{2})$. 
In Fig.~\ref{fig:Energy}(a, b) we show the energy as a function of the number of full imaginary time sweeps for the ITF model at field strength $\lambda=3.5$ compared to a near-exact DMRG result. Note that both $D$ (which controls the variational space
of the standard PEPS) and $D_c$ (vertical bond dimension of the central column, which controls the accuracy of
the absorption step in the canonicalization sweep) affect the final converged energy; in this setting, increasing $D_c$ has a larger effect than increasing $D$. The relative error of the energy per site reached for the largest bond dimension $D=4, D_c=8$ for both $4\times 4$ and $8\times 8$ lattice sizes is on the order of $10^{-4}$. 

{In Fig.~\ref{fig:Energy}(c, d), we benchmark the variational energy for the canonical PEPS and the standard PEPS for the ITF model at $\lambda=3.5$ and the Heisenberg model. For the standard PEPS we used the optimization algorithm from Ref.~\onlinecite{Lubasch:2014} to obtain the ground state. Both algorithms are initialized by the same random PEPS. The maximum relative errors of the QR and absorption steps using the two layer, $l_y=2$ QR ansatz are on the order of $10^{-4}$ and $10^{-3}$ respectively in the ITF and Heisenberg simulation. This relative error bounds the ultimate accuracy in the energy that the canonical PEPS simulation can achieve. Thus we observe that standard PEPS optimization leads to lower energies than the canonical PEPS energy in both cases as expected. }

\section{Conclusions}
\label{Sec:Conclusion}

In conclusion, we have described a procedure to convert a PEPS  into a canonical form analogous to that of an MPS where all columns but one are isometric, by sequentially decomposing columns through a variational QR ansatz. We find that the canonicalization
is stable and can be carried out with a small and controllable error. Canonicalization redistributes entanglement in the PEPS, resulting in a central column with increased bond dimension.
%% The cost of the algorithms implemented in the canonical form thus depends both on the intrinsic bond dimension and the
%% increased central column bond dimension.
Our procedure introduces the possibility to formulate canonical PEPS algorithms which make explicit use of an isometric
environment, which we demonstrated in an imaginary time optimization of the ground-state energy.
{The canonical form is clearly numerically more stable, as we show in calculations of the
  stability of the norm with respect to perturbations of the tensors.
However, a faithful comparison of the cost of algorithms using the canonical PEPS and standard PEPS requires considerably more analysis. This is because the canonicalization leads to a non-homogeneous PEPS with different bond dimensions on the vertical and horizontal bonds, quite different from the standard PEPS scenario, and in addition, the canonical PEPS introduces additional numerical parameters (to control the accuracy of the QR ansatz and $R$ absorption) which must be converged. To eliminate the additional errors of the QR decomposition, it may be more expedient to directly optimize the underlying network of isometries and unitaries suggested by the QR ansatz, without explicitly converting into the standard PEPS column-row form. Future investigations will focus on more detailed analysis of these and other algorithms as well as the general representational power of canonicalized PEPS.}

\textit{Note}: As this manuscript was prepared for submission, we were notified of Ref.~\onlinecite{zaletel2019isometric}. In that work, the authors similarly
pursue a full canonicalization of a PEPS, but use a different set of isometric conditions on the $Q$ tensors which are more constrained than the ones that we use. Further work is needed to understand the relationship between these techniques.

\begin{acknowledgements}
Primary support for this work was from MURI FA9550-18-1-0095.
Some of code used to test energy optimization was
based on work supported by the US National Science Foundation (NSF) via grant CHE-1665333.
MJO acknowledges a US NSF Graduate Research Fellowship via grant
DEG-1745301. GKC acknowledges support from the Simons Foundation. We have used \emph{Uni10} \cite{Kao:2015} as a middleware library to build up variational gauging ansatz.

\end{acknowledgements}

\bibliography{Ref}

\newpage
\onecolumngrid

\end{document}